\documentstyle[twoside,fleqn,espcrc2]{article}

\def\ypbco#1 #2{$\rm Y_{#1}Pr_{#2}BCO$}
\def\YPBCO#1 #2{$\rm Y_{#1}Pr_{#2}Ba_2Cu_3O_7$}

\newcommand{\AmS}{{\protect\the\textfont2
  A\kern-.1667em\lower.5ex\hbox{M}\kern-.125emS}}

\title{Complex conductivity of $\rm Y_{1-x}Pr_xBa_2Cu_3O_7$ thin
       films measured by coherent terahertz spectroscopy}

\author{
J.O. White\thanks{on leave from: Hughes Research Laboratories, Malibu.},
R. Buhleier,
S.D. Brorson\thanks{present addr.: Teledanmark Laboratories, Copenhagen.},
I.E. Trofimov\thanks{present address: Rutgers University, Piscataway, NJ.},
H.-U. Habermeier, and J. Kuhl
\address{Max-Planck-Institut f\"ur Festk\"orperforschung,
        Postfach 800665, 70506 Stuttgart, Germany}
}
\begin{document}

\begin{abstract}
The complex transmission of $\rm Y_{1-x}Pr_xBa_2Cu_3O_7$ single crystal
thin films has been measured in the range 0.2-1.0 THz using time domain
spectroscopy.
The complex conductivity is calculated without using a Kramers-Kronig
analysis.
All of the superconducting samples
show a peak in $\sigma_1(T)$ below $T_c$.
The underdoped samples show a deviation from $1/(\alpha+\beta T)$ behavior
above $T_c$ that may be linked with the onset of a spin gap.
\end{abstract}

\maketitle

\section{INTRODUCTION}
The spectroscopic data on the high-$T_c$
superconducting compound $\rm Y_1Ba_2Cu_3O_7$
measured in the far infrared (FIR) as well as the GHz-region
has revealed many interesting features.
The new technique of THz-time-domain spectroscopy bridges the
gap between the two regimes.
Here we report results of a THz investigation into the
\YPBCO {1-x} x\(YPrBCO) system.
We investigated five YPrBCO samples having Pr composition
$x$ = 0.0, 0.2, 0.3, 0.4, and 1.0.
The films are grown to a thickness of
150~nm by pulsed laser deposition \cite{habermeier91} onto
NdGaO$_3$ substrates.

Our spectroscopic technique involves a time-domain measurement
of the electric field of a
microwave pulse transmitted through the sample \cite{brorson94}.
A Fourier transform yields the complex transmission spectrum.
To calculate the conductivity, we make use of a multiple reflection
formula for the field transmitted through the YBCO layer.

\section{RESULTS}
The conductivity spectra at 50~K are shown in Fig.~\ref{sigma_nu}.
The addition of Pr to YBCO should have at least two interrelated effects:
a) The suppression of $T_c$ changes the partitioning between normal and
superconducting carriers.  b) The total number of
carriers $N$ (or their mobility) may be reduced.
\begin{figure}[htb] 
\makebox[55mm]{\rule[-21mm]{0mm}{75mm}}
\caption{$\sigma$ vs. $\nu$ for \YPBCO{1-x} x\ at $T=50$~K.
The solid and dashed upper curves are fits to a Drude model
for $\sigma_1$ and to $1/\nu$ for $\sigma_2$.}
\label{sigma_nu}
\end{figure}
Both factors a) and b) cause $\sigma_2$ to decrease with [Pr].
Only normal carriers contribute to $\sigma_1$ for $\omega \neq 0$,
but now factors a) and b) compete.
At 50~K, $\sigma_1$ decreases with [Pr], therefore
the effect of a reduction in $N$ dominates the
effect of the shift in $T_c$.
For 30\% and 40\% Pr, we observe a frequency dependent $\sigma_1$
and can directly measure $\tau(T)$, the quasi-particle scattering time.
Pure PrBCO is a dielectric at 50~K, as seen by a conductivity
proportional to frequency.

Examining $\sigma_2(T)$ at a fixed frequency (Fig.~\ref{sigma_T}a),
we see that it is close to zero at high temperature,
but rises sharply at the onset of superconductivity,
thus providing an ac measurement of $T_c$ (Table 1).
\begin{figure}[tbp] 
\makebox[55mm]{\rule[-21mm]{0mm}{90mm}}
\caption{$\sigma$ vs. T at 480~GHz
for the four Pr concentrations $x = 0$ ($\Box$),
0.2 ($\circ$), 0.3 ($\bigtriangleup$),
0.4 ($\bigtriangledown$).
The vertical lines indicate $T_c^{ac}$.
For clarity, the data for $x=0.3$ in b) has been multiplied by 1.5.
}
\label{sigma_T}
\end{figure}
In all of the superconducting alloys, $\sigma_1$ displays
a peak below $T_c$ which has been seen previously only in pure YBCO.
It has been attributed to a sharp rise in the scattering time
dominating the effect of a decrease in the number of
normal carriers \cite{nuss91}.

The normal state behavior of our samples is particularly interesting
because other {\em underdoped} materials such as
oxygen deprived (123)YBCO undergo
a phase transition associated with the opening of a spin gap
at $T_D>T_c$ \cite{ito93}.
If the normal carriers couple strongly to spin
fluctuations, the opening of a spin gap should be accompanied by an
{\em increase} in the scattering time $\tau$, giving rise to an
{\em enhancement} in $\sigma_1$ below $T_D$ for $\omega<1/\tau$.

For pure (optimally doped) YBCO, at 480~GHz, $\sigma_1$ shows only a
single transition at $T_c$.
For the (underdoped) alloys,
$\sigma_1$ has two transitions, one at $T_c$, the other at a
higher temperature which increases with [Pr].
To accentuate the two transitions, Fig.~\ref{sigma_T}b is shaded in the
region bounded by $T_c$, the experimental curve, and a dashed line
representing $1 / (\alpha+\beta T)$ behavior.
The higher transition temperature seen at 480~GHz matches that of a
transition also observed in the dc resistivity.

We evaluate the penetration depth $\lambda_L$
and the plasma frequency $\omega_p$ by fitting the data
to a two fluid model of the form:
\begin{equation}
\sigma(\omega) = {\epsilon_0 \omega_p^2 \tau
\over {1-i\omega\tau}}x_n +
{1 \over {\mu_0\lambda_L^2}}\left(- \pi\delta(\omega) + {i \over \omega}
\right) x_s,
\label{2fluid}
\end{equation}
where $x_n$ and $x_s$ are the fractions of normal and
superconducting carriers.
The results are shown in Table 1.
The decrease in $\omega_p$ with [Pr] supports the theory that
Pr suppresses superconductivity by reducing the population of
mobile holes in the $\rm CuO_2$ planes.
\begin{table}[hbt]
\caption{}
\begin{tabular}{|c||c|c|c|c|} 
\hline
$x$  & $d$ (nm) & $\lambda_L$ (nm) & $\omega_p$ (cm$^{-1}$)
& $T_c^{\rm ac}$ (K) \\
\hline \hline
0.0 & 155 & 170 & 9500  & 92  \\ \hline
0.2 & 134 & 350 & 4600 & 72 \\ \hline
0.3 & 170 & 380 & 4200 & 59 \\ \hline
0.4 & 170 & 590 & 2700 & 41 \\
\hline
\end{tabular}
\end{table}

\begin{thebibliography}{9}

\bibitem{habermeier91}
H.~U. Habermeier {\it et~al.}, Physica C {\bf 180},  17  (1991).

\bibitem{brorson94}
S.~D. Brorson {\it et~al.}, Phys. Rev. B {\bf 49},  6185  (1994).

\bibitem{nuss91}
M.~C. Nuss {\it et~al.}, Phys. Rev. Lett {\bf 66},  3305  (1991).

\bibitem{ito93}
T. Ito, K. Takenaka, and S. Uchida, Phys. Rev. Lett. {\bf 70}, 3995 (1993).

\end{thebibliography}
\end{document}